\documentclass[aps,amsmath,amssymb,pr,nofootinbib,twocolumn]{revtex4-2}
\usepackage{bm}
\usepackage{epsfig}
\usepackage{amsmath}
\usepackage{color}
\newcommand{\nix}[1]{}
\usepackage[unicode=true,colorlinks=true,citecolor=blue]{hyperref}
\usepackage{physics}

\definecolor{darkviolet}{rgb}{0.58, 0.0, 0.83}

\begin{document}

\title{Weak localization in $\bf p$-type heterostructures in the presence of parallel magnetic field}

\author{M.\,O.\,Nestoklon and L.\,E.\,Golub}
\affiliation{Ioffe Institute, 194021 St.~Petersburg, Russia}

\begin{abstract}
Theory of weak localization is developed for two-dimensional holes in the presence of in-plane magnetic field. The Zeeman splitting even in the hole momentum results in the spin-dependent phase changing the quantum interference. The negative correction to the conductivity is shown to decrease by a factor of two by the in-plane magnetic field. The positive magnetoconductivity in a classically weak perpendicular field caused by the weak localization is calculated for both quadratic and quartic in momentum Zeeman hole splittings. Calculations show that the conductivity corrections are very close to each other in these two cases of low and high hole density.
\end{abstract}

\maketitle

\section{Introduction}

Spin-dependent phenomena attract a great attention due to spin-orbit interaction allowing spin manipulation by electrical or optical means. The first important move in this way was a discovery of the Rashba splitting of electron energy spectrum in bulk wurtzite-type semiconductors~\cite{Rashba_1960}.
In two-dimensional (2D) systems, this splitting is present in heterostructures made of any material provided the structure inversion asymmetry is present~\cite{Golub_Ganichev_pss}.
Generally, the spin-orbit interaction is described by the term in the Hamiltonian which can be presented in the form
\begin{equation}
\label{H}
\mathcal H_\text{SO} = \hbar \bm \sigma \cdot \bm \Omega,
\end{equation}
where 
$\bm \sigma = (\sigma_x, \sigma_y)$ is a vector of Pauli matrices, and the spin-orbit splitting equals to $2\hbar\Omega$.

The spin-orbit splitting of the electron energy spectrum leads to many interesting optical and transport  phenomena~\cite{Dyakonov_book}. In transport, it leads to a remarkable beating patterns in the Shubnikov-de~Haas oscillations where it can be easily detected. However, a good mobility is needed for such kind of manifestation of the Rashba splitting which should be much larger than the level broadening. 
Nevertheless, even in low-mobility samples the Rashba splitting can be measured. This can be done in classically-low magnetic fields where the magnetoresistance is caused by the weak localization (WL) effect, see Ref.~\cite{SST_MG_LG} for review. Developed theoretical expressions for the WL correction to the conductivity valid for arbitrary values of the Rashba splitting allow adequately extracting the splitting value and other electron kinetic and band parameters by fitting the experimental data.

2D holes in semiconductor heterostructures represent a system which is very different from electrons. This happens because the holes in the ground 2D subband have spin projection $\pm 3/2$ on the structure main axis. In particular, they have a cubic in momentum Rashba splitting~\cite{Winkler_book}. Due to the same reason, the Zeeman splitting of heavy-holes in the in-plane magnetic field 
at the bottom of the 2D subband is cubic in the field strength in the axial approximation. A small contribution for free holes is present due to cubic symmetry of the zinc-blende lattice forming the heterostructure~\cite{beats_1998} which, however, increases strongly for localized holes in quantum dots~\cite{Trifonov_2021}.
At finite wavevectors the situation changes, and the momentum-dependent in-plane Zeeman splitting arises. In the axial approximation, the Hamiltonian of heavy holes in the ground subband of a symmetrical quantum well in the presence of an in-plane magnetic field $\bm B_\parallel$ is given by~\cite{Miserev_Sushkov}
\begin{equation}
\mathcal H_\text{SO} = \hbar \sigma_-(\Delta_1 B_+k_+^2 +  \Delta_2 B_-k_+^4) + h.c.
\end{equation}
Here 
$\bm k$ is the in-plane wavevector, 
$\sigma_\pm=(\sigma_x\pm i\sigma_y)/2$ with the operators $\sigma_{x,y}$ coupling two Kramers-degenerate hole states, $B_\pm=B_x\pm iB_y$, $k_\pm=k_x\pm ik_y$, and $\Delta_{1,2}$ are constants.
This expression coinsides with Eq.~\eqref{H} where  $\bm \Omega$ is given by
\begin{equation}
\label{Omega}
\Omega_x + i\Omega_y = B_\parallel \left( \Delta_1 k^2 \text{e}^{2i\varphi} + \Delta_2 k^4 \text{e}^{4i\varphi} \right)
\end{equation}
with $\varphi$ being an angle between $\bm k$ and $\bm B_\parallel$.

According to estimates given in Ref.~\cite{marcelina}, the Zeeman splitting at $B_\parallel =1$~T is  $2\hbar \Omega \sim 0.1 \ldots 1$~meV. This allows considering the Zeeman splitting being much smaller than the Fermi energy and  affecting WL as an additional phase in the electron interference. 
This allows us to solve the WL problem by the method used in Refs.~\cite{SST_MG_LG,Low_symm_QW_2013} assuming the ratio of the splitting and level broadening to be arbitrary but ignoring the difference in the Fermi wavevectors in spin subbands.

We study two limits 
of low and high hole densities where $\Delta_1 k_\text{F}^2$ prevails over $\Delta_2 k_\text{F}^4$ or vice versa. Here $k_\text{F}$ is the Fermi wavevector.
In both cases, the Zeeman splitting is isotropic in the $\bm k$-space, and the WL problem can be solved analytically. The hole Hamiltonian~\eqref{H} is \emph{even} in $\bm k$, therefore the WL correction to conductivity   and the anomalous magnetoresistance are negative.
We consider diffusive and ballistic regimes of WL~\cite{DKG_1997} where the interference contribution to the conductivity occurs on large and small trajectories, respectively.

\section{WL conductivity correction in diffusive regime}

In the low-density limit, the Hamiltonian~\eqref{H} with $\bm \Omega$ from Eq.~\eqref{Omega} has the same form as that of exciton-polaritons in microcavities with $\Omega=\Delta_1 k_\text{F}^2 B_\parallel$ 
instead of the longitudinal-transverse 
splitting, see Ref.~\cite{polaritons} for details. 
Therefore the WL correction to conductivity $\sigma <0$ in the diffusion approximation is given by the expression following from Ref.~\cite{polaritons}:
\begin{equation}
\label{ds0}
\Delta\sigma(0) = -{e^2\over 4\pi^2\hbar} \left[ 2\ln{\left(T_{1}\over \tau \right)} - \ln{\left({T_{0}\over\tau} \right)} + \ln{\left(\tau_\phi\over \tau \right)}\right] .
\end{equation}
Here $\tau$ is the transport scattering time, $\tau_\phi$ is the dephasing time, and the spin relaxation rates are given by~\cite{footnote1}
\begin{equation}
{1\over T_1} = {1\over \tau_s} + {1\over \tau_\phi}, \qquad 
{1\over T_0} = {2\over \tau_s} + {1\over \tau_\phi},
\qquad
{1\over \tau_s} = 2\Omega^2\tau.
\end{equation}

The first two terms in Eq.~\eqref{ds0} is due to interference in the triplet channel while the last one is caused by the singlet channel. It is worth mentioning that the classification on triplet and singlet here is by the difference of angular momenta of two interfering particle waves rather than by their sum. Furthermore, the sign of the $\Omega$-independent singlet contribution is positive while one of triplet terms is negative. This difference with a traditional results for spin-orbit coupled systems is due to even in $\bm k$ spin-orbit splitting, cf. Ref.~\cite{polaritons}.
%

In the presence of a small perpendicular magnetic field $\bm B \parallel z$, 
we are interested in the magnetoconductivity $\delta\sigma=\sigma(B_z)-\sigma(B_z=0)$. 
In the diffusion approximation $B_z \ll B_\text{tr}$ where the transport field is defined as
\begin{equation}
B_\text{tr}={\hbar \over 2\abs{e}l^2}
\end{equation}
with $l$ being the mean free path, we have
\begin{equation}
\label{ds_diff}
\delta\sigma = {e^2\over 4\pi^2\hbar} \left[ 2f_2{\left({B_z\over B_{1}} \right)} - f_2{\left({B_z\over B_{0}} \right)} + f_2\left({B_z\over B_\phi} \right)\right].
\end{equation}
Here $f_2(x)=\psi(1/2+1/x)+\ln{x}$ with $\psi$ being the digamma-function,
$B_\phi=B_\text{tr}\tau/\tau_\phi$
and $B_{1,0}=B_\phi\tau_\phi/T_{1,0}$.

At $\Omega\tau = 0$, the correction is $\delta\sigma^0(B_z) = {e^2\over 2\pi^2\hbar}f_2\left({B_z/ B_\phi} \right)$.
At large $\Omega\tau \lesssim 1$, the first two terms are much smaller than the last one. 
As a result, the correction is twice smaller than at $\Omega=0$: $\delta\sigma^\infty(B_z) = {e^2\over 4\pi^2\hbar}f_2\left({B_z/ B_\phi} \right)$.

At high density the results for both zero-field correction and the magnetoconductivity are the same, the only difference is that $\Omega=\Delta_2 k_\text{F}^4 B_\parallel$.
Differences in the functional forms of the WL contribution to the conductivity at low and high densities appear 
in stronger perpendicular fields $B_z \sim B_\text{tr}$. In this case the ballistic trajectories with a few, three or more, impurities contribute to the conductivity, therefore this is called \emph{ballistic regime} of WL.

\section{Ballistic regime}

Here we also take into account non-logarithmic corrections to the conductivity as well as the non-backscattering contribution. 
We search for the the Cooperon -- the interference correction to the probability density for an electron to reach the point $\bm r'$ starting from the point $\bm r$.
The Cooperon 
satisfies the integral equation which can be presented in the following form~\cite{WL_spin_nondiff}:
\begin{equation}
\mathcal C(\bm r, \bm r') = \mathcal P(\bm r, \bm r') + \int \dd \bm r'' \mathcal P(\bm r, \bm r'')\mathcal C(\bm r'', \bm r').
\end{equation}
Here the kernel 
is given by
\begin{equation}
\mathcal P(\bm r, \bm r') = \mathcal P_0(R)\exp[2i\varphi(\bm r, \bm r')]\exp[-2i\tau \bm L \cdot \bm \omega(\bm R)],
\end{equation}
where $\bm R = \bm r - \bm r'$,  $\mathcal P_0(R)=\exp(-R/\tilde{l})/(2\pi Rl)$ with $l$ being the mean-free path,  $\tilde{l}=l/(1+\tau/\tau_\phi)$, the magnetic phase $\varphi(\bm r, \bm r') = (x+x')(y'-y)/(2l_B^2)$ with $l_B=\sqrt{\hbar c/\abs{eB_z}}$ being the magnetic length of the unitary charge, and $\bm L$ is an operator of a difference of spins of two interfering particles.

\subsection{$\bf k^2$-splitting}

First we consider the $k^2$-term in the spin-splitting with $\Omega_+=\Omega \exp(2i\varphi)$, $\Omega=\Delta_1 k_\text{F}^2 B_\parallel$.
In this case, 
the vector $\bm \omega$ is given by
\begin{equation}
\label{omega_r}
\omega_\pm =  \omega n_\mp^2, \qquad \omega= \Omega {R\over l }, \qquad \bm n = \bm R/R.
\end{equation}
The operator $\exp[-2i\tau \bm L \cdot \bm \omega(\bm R)]$
implies that it is block-diagonal in the double-charge Landau-level basis where the levels 
$N$, $N-2$ and $N-4$ are mixed. 
In this basis the triplet part of the operator $\mathcal P$ can be presented in the following form:
\begin{equation}
\label{AN}
A_N = 
\begin{bmatrix}
P_N - S_N^{(0)}& -i R_{N}^{(2)}&S_{N}^{(4)}\\
-i R_{N}^{(2)}&P_{N-2} - 2S_{N-2}^{(0)}&-i R_{N-2}^{(2)}\\
S_{N}^{(4)}&-i R_{N-2}^{(2)}&P_{N-4} - S_{N-4}^{(0)}
\end{bmatrix}.
\end{equation}
Here 
\begin{equation}
\label{PN}
P_N = {l_B\over l}\int\limits_0^\infty \dd x \exp\qty(-x{l_B\over \tilde{l}} - {x^2\over 2})L_N(x^2)
, 
\end{equation}
\begin{align}
\label{RN}
& R_{N}^{(m)} = {l_B\over l\sqrt{2}}\sqrt{(N-m)!\over N!}\\
& \times \int\limits_0^\infty \dd x \exp\qty(-x{l_B\over l} - {x^2\over 2})x^m L_{N-m}^{(m)}(x^2)\sin\qty(2\Omega\tau{l_B\over l}x)
, \nonumber
\end{align}
\begin{align}
\label{SN}
& S_{N}^{(m)} = {l_B\over l}\sqrt{(N-m)!\over N!} \\
& \times \int\limits_0^\infty \dd x \exp\qty(-x{l_B\over l} - {x^2\over 2})x^m L_{N-m}^{(m)}(x^2)\sin^2\qty(\Omega\tau{l_B\over l}x), \nonumber
\end{align}
and we assume $P_N$ with $N<0$ and $R_{N}^{(m)}$, $S_{N}^{(m)}$ with $N<m$ equal to zero.

The backscattering contribution to the conductivity is given by
\begin{equation}
\label{sigma_bs}
\sigma_{\rm bs} = -{e^2 \over 2\pi^2\hbar} \qty(l\over l_B)^2\sum_{N=0}^\infty \left[\text{Tr}\qty(\Pi A_N^2 \mathcal C_N) + {P_N^3\over 1-P_N} \right],
\end{equation}
where $\Pi = \text{diag}(1,-1,1)$, and the 
Cooperon 
matrix $\mathcal C_N$ reads
\begin{equation}
\mathcal C_N = A_N(I-A_N)^{-1}
\end{equation}
with $I$ being the $3\times 3$ unit matrix.

The non-backscattering contribution reads~\cite{WL_spin_nondiff}
\begin{equation}
\label{sigma_nbs}
\sigma_{\rm nbs} = -{e^2 \over 4\pi^2 \hbar} \qty(l\over l_B)^2 
\text{Tr}\biggl[\Pi \biggl(
\mathcal J^+ \mathcal C \mathcal J^-  +\mathcal  J^- \mathcal C  \mathcal J^+
\biggr) \biggr],
\end{equation}
where 
$\mathcal J^\pm = n_{\pm}\mathcal P$ are the vertex operators.
Calculation yields
\begin{align}
\label{sigma_nbs1}
\sigma_{\rm nbs} = & {e^2 \over 4\pi^2\hbar} \qty(l\over l_B)^2 \sum_{N=0}^\infty \biggl\{ \qty(Q_{N+1}^2+ Q_{N}^2) {P_N\over 1-P_N}\nonumber\\
&+ \text{Tr}
\qty[\Pi (K_{N+1}^T K_{N+1} +  K_{N} K_{N}^T)  {\mathcal C}_N ] 
\biggr\}. 
\end{align}
Here 
\begin{equation}
Q_N = {l_B\over l}{\Theta(N)\over \sqrt{N}}\int\limits_0^\infty \dd x \exp\qty(-x{l_B\over {l}} - {x^2\over 2}) x L_{N-1}^{(1)}(x^2),
\end{equation}
and
\begin{equation}
K_{N}=
\begin{bmatrix}
Q_{N} - S_{N}^{(1)}&  -i R_{N}^{(3)}& S_{N}^{(5)}\\
i  R_{N-1}^{(1)}&Q_{N-2} - 2S_{N-2}^{(1)}& -i R_{N-2}^{(3)}\\
- S_{N-1}^{(3)}&i R_{N-3}^{(1)}&Q_{N-4} - S_{N-4}^{(1)}
\end{bmatrix}.
\end{equation}

\subsection{Zero perpendicular field}

The limit $B_z \to 0$ is evaluated in a standard way. In the absense of spin-orbit splitting, $\Omega=0$, we have a well-known result
\begin{equation}
\label{ds0_nondiff}
\Delta\sigma(0) = -{e^2\over 2\pi^2\hbar} \ln({\tau_\phi\over 2\tau}).
\end{equation}
Here the two contributions are given by $\sigma_{\rm bs}(0) =-{e^2\over 2\pi^2\hbar} \ln({\tau_\phi/ \tau})$ and $\sigma_{\rm nbs}(0) ={e^2\over 2\pi^2\hbar} \ln2$, cf. Eq.~\eqref{ds0}.

At arbitrary values of the product $\Omega\tau$, the zero-$B_z$ correction is obtained  by  passing in Eqs.~\eqref{sigma_bs}, \eqref{sigma_nbs1} from summation to integration~\cite{MN_NS_ST_graphene_nondiff}:
\begin{equation}
\label{subst_N_x}
\qty(l\over l_B)^2\sum_{N=0}^\infty \to {1\over 4}\int\limits_0^\infty \dd x, \qquad x = {4N}\qty(l\over l_B)^2.
\end{equation}
The values 
$P_N$, $Q_N$, $R_{N}^{(m)}$ and $S_{N}^{(m)}$ are changed to $P(x)$, $Q(x)$, $R_{m}(x)$ and $S_{m}(x)$,
and  matrices $A_N$ and $K_N$ are changed to $A_x$ and $K_x$ accordingly. Using the asymptotics of Laguerre polynomials at $N \to \infty$ we obtain 
\begin{align}
\label{Px}
& P=  {1\over \sqrt{(l/\tilde{l})^2+x}}, 
\qquad 
Q=P \sqrt{1-P\over 1+P},
\\
& R_{m}={\text{Im} T_m\over \sqrt{2}} ,
\quad
 S_{m}=
{1\over 2}\qty[P\qty({1-P\over 1+P})^{m/2} - \text{Re}T_m], \nonumber
\end{align}
where
\begin{align}
&T_0= {P \over \sqrt{1-4\Omega\tau(\Omega\tau+i)P^2}},
\\
&T_m = T_0 \qty[{\sqrt{1-4\Omega\tau(\Omega\tau+i)P^2}+(2i\Omega\tau-1)P\over \sqrt{1-P^2}}]^{m}. \nonumber
\end{align}

It is convenient then to change the integration variable to $P$:
\begin{equation}
\label{zero_field_sigma_bs}
\sigma_{\rm bs}(0) = -{e^2 \over 4\pi^2\hbar} \left[ \int\limits_0^{\tilde{l}/l} \dd P {\text{Tr}\qty(\Pi A_x^2\mathcal C_x)\over  P^3} + \ln({\tau_\phi \over \tau})\right],
\end{equation}
\begin{align}
\label{zero_field_sigma_nbs}
\sigma_{\rm nbs}(0) = {e^2 \over 8\pi^2\hbar}\int\limits_0^1 \dd P & {\text{Tr}\qty[\Pi (K_x^T K_x +  K_{x} K_{x}^T) {\mathcal C}_x ] \over P^3}  \nonumber \\
&+ {e^2\over 4\pi^2\hbar} \ln2.
\end{align}
Here
$\mathcal C_x = A_x(I-A_x)^{-1}$. We keep the difference between $\tilde{l}$ and $l$  only in~\eqref{zero_field_sigma_bs}  because it is important for cutting the pole in the integrand.

In the limit $\Omega\tau \to \infty$ we have $R_{m}(x)=0$. This means  the interference 
in one channel is totally suppressed by spin-orbit interaction. Numerical integration with the 2-rank matrices $A_x$, $K_x$ yields in this limit
\begin{equation}
\label{ds0_Omega_infty}
\sigma_{\rm bs}^\infty(0) =-{e^2 \over 4\pi^2\hbar} \qty(\ln{\tau_\phi\over \tau} + 0.4514),
\quad
\sigma_{\rm nbs}^\infty(0) 
= 0.209{e^2 \over \pi^2\hbar}.
\end{equation}

\subsection{$\bf k^4$-splitting}

At high hole density, when the $B_\parallel k^4$-term dominates in the Zeeman splitting, the vector $\bm \omega(\bm R)$ introduced in Eq.~\eqref{omega_r} changes to
\begin{equation}
\omega_\pm = \omega n_\mp^4, \qquad \Omega=\Delta_2 k_\text{F}^4 B_\parallel.
\end{equation}
This results in the following matrix $A_N$ instead of Eq.~\eqref{AN}
\begin{equation}
\label{AN_k4}
A_N = 
\begin{bmatrix}
P_N - S_N^{(0)}& -i R_{N}^{(4)}&S_{N}^{(8)}\\
-i R_{N}^{(4)}&P_{N-4} - 2S_{N-4}^{(0)}& -i R_{N-4}^{(4)}\\
S_{N}^{(8)}& -i R_{N-4}^{(4)}&P_{N-8} - S_{N-8}^{(0)}
\end{bmatrix}.
\end{equation}
The backscattering contribution to the conductivity is given by Eq.~\eqref{sigma_bs}.

The nonbackscattering correction $\sigma_\text{nbs}$ is given by the general Eq.~\eqref{sigma_nbs}.
In the case of splitting $\propto k^4$,
we again obtain Eq.~\eqref{sigma_nbs1} with the matrix $K_N$ given by
\begin{equation}
K_{N} = 
\begin{bmatrix}
Q_{N} - S_{N}^{(1)}&  -i R_{N}^{(5)}& S_{N}^{(9)}\\
i  R_{N-1}^{(3)}&Q_{N-4} - 2S_{N-4}^{(1)}& -i R_{N-4}^{(5)}\\
- S_{N-1}^{(7)}& i R_{N-5}^{(3)}& Q_{N-8} - S_{N-8}^{(1)}
\end{bmatrix}.
\end{equation}

At $B_z=0$, the conductivity corrections are also given by Eqs.~\eqref{zero_field_sigma_bs},~\eqref{zero_field_sigma_nbs} where the matrices $A_x$ and $K_x$ are obtained from the above given matrices $A_N$, $K_N$ by the substitutions~\eqref{Px}.

In the limit $\Omega\tau \to \infty$ we have
\begin{equation}
\label{ds0_Omega_infty_k4}
\sigma_{\rm bs}^\infty(0) =-{e^2 \over 4\pi^2\hbar} \qty(\ln{\tau_\phi\over \tau} + 0.3964),
\quad
\sigma_{\rm nbs}^\infty(0) 
= 0.2059{e^2 \over \pi^2\hbar}.
\end{equation}

\section{Results and Discussion}

For numerical calculations of the conductivity corrections, we extend the approach used in Ref.~\cite{MN_NS_mcond_intervalley}. For the computation of the sums~\eqref{sigma_bs},~\eqref{sigma_nbs1} we use the following procedure: we choose a value of a high Landau level $N_{max}$. 
Next, the infinite sums ~\eqref{sigma_bs},~\eqref{sigma_nbs1} are split in two parts: for $N\leq N_{max}$ we directly sum the terms, and for $N>N_{max}$ we replace the sums with the integral of the matrix function obtained from Eqs.~\eqref{sigma_bs},~\eqref{sigma_nbs1} by using the large-$N$ limit \eqref{Px}. 
In fact, this part of the correction is given by Eqs.~\eqref{zero_field_sigma_bs},~\eqref{zero_field_sigma_nbs} with the dependence on magnetic field in the upper integral limit.
For the results below we find $N_{max}=300$ sufficient for converging up to fourth digit in the whole range of magnetic fields. For small and large values of $B_z$, one may use smaller $N_{max}$ as in the former case the correction is defined by large Landau levels and in the latter case the correction is defined by small number of Landau levels close to zero. For small magnetic fields, the evaluation of integrals in \eqref{PN} is impractical as the function in the integral is oscillating for large $N$. In this case, we use the reccurence relations given in Appendix~\ref{sec:rec_rel}. Recurrence relations are used for $\mathcal P_N^{(0)}$ and $\mathcal P_N^{(1)}$, but for $2\le m\le 4$ (large $B_z$ and/or $\Omega\tau$) and $m \ge 5$ one has to use the direct calculation of the integrals \eqref{PN}. As a result, the calculation of the conductivity correction for $k^4$-splitting is more time consuming but still feasible.

\begin{figure}[t]
\centering
\includegraphics[width=0.95\linewidth]{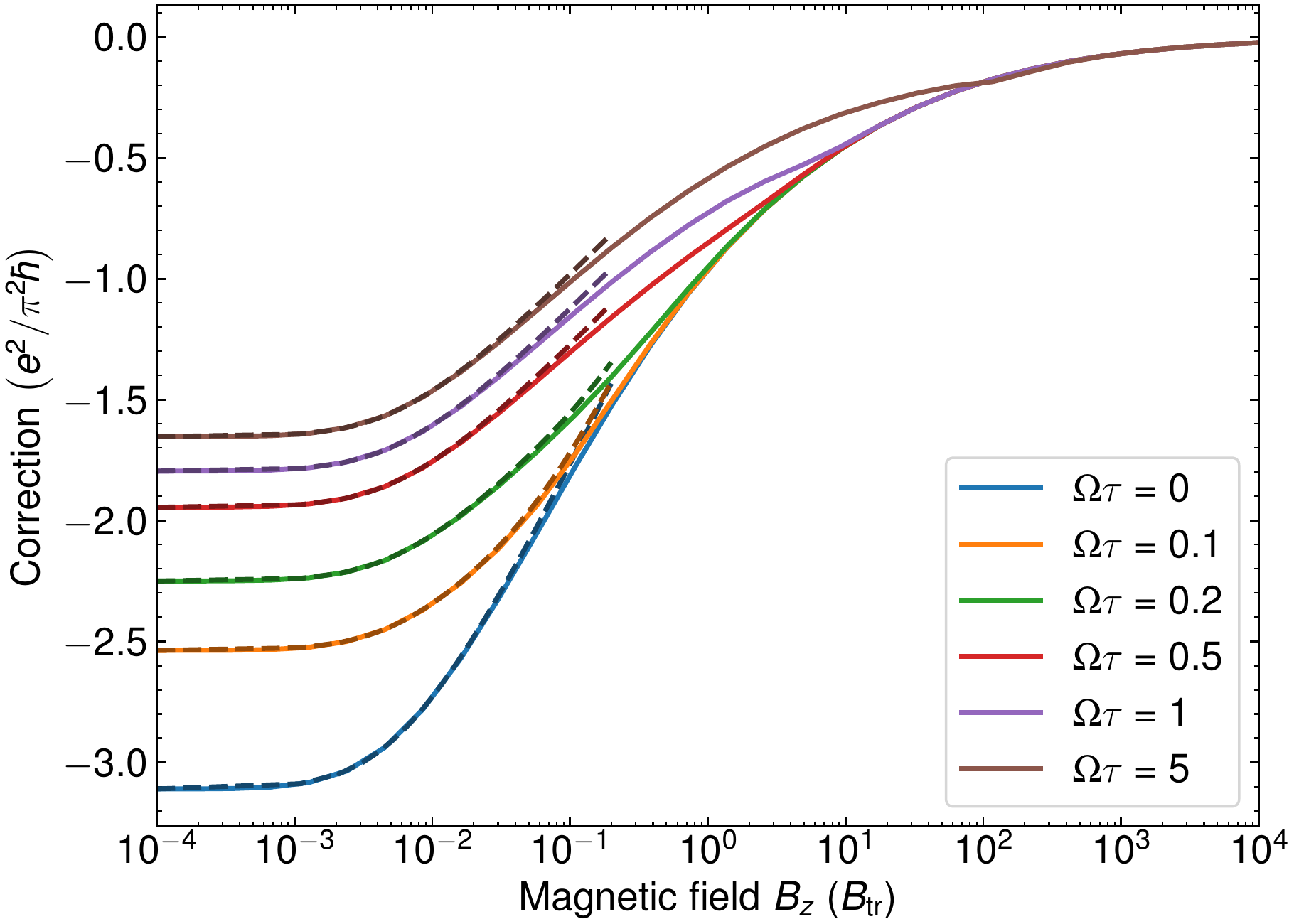}
\caption{Conductivity correction at  $k^2$-splitting as a function of $B_z/B_\text{tr}$ for various $\Omega\tau \propto B_\parallel$. The dephasing time $\tau_\phi/\tau=10^3$. 
Diffusion approximation results~\eqref{ds_diff} are shown by dashed lines.}
\label{fig_mcond}
\end{figure}

In Fig.~\ref{fig_mcond}, the conductivity correction is shown as a function of the $B_z$ for various values of the spin-orbit splitting $\Omega \propto B_\parallel k^2$. The zero-$B_z$ value at large $\Omega\tau$ is twice smaller than at $\Omega=0$. At large $B_z \gg B_\text{tr}$ all curves tend to the same dependence because of absence of spin rotations at characteristic trajectories with the size $\sim l_B \ll l$.

\begin{figure}[b]
\centering
\includegraphics[width=0.95\linewidth]{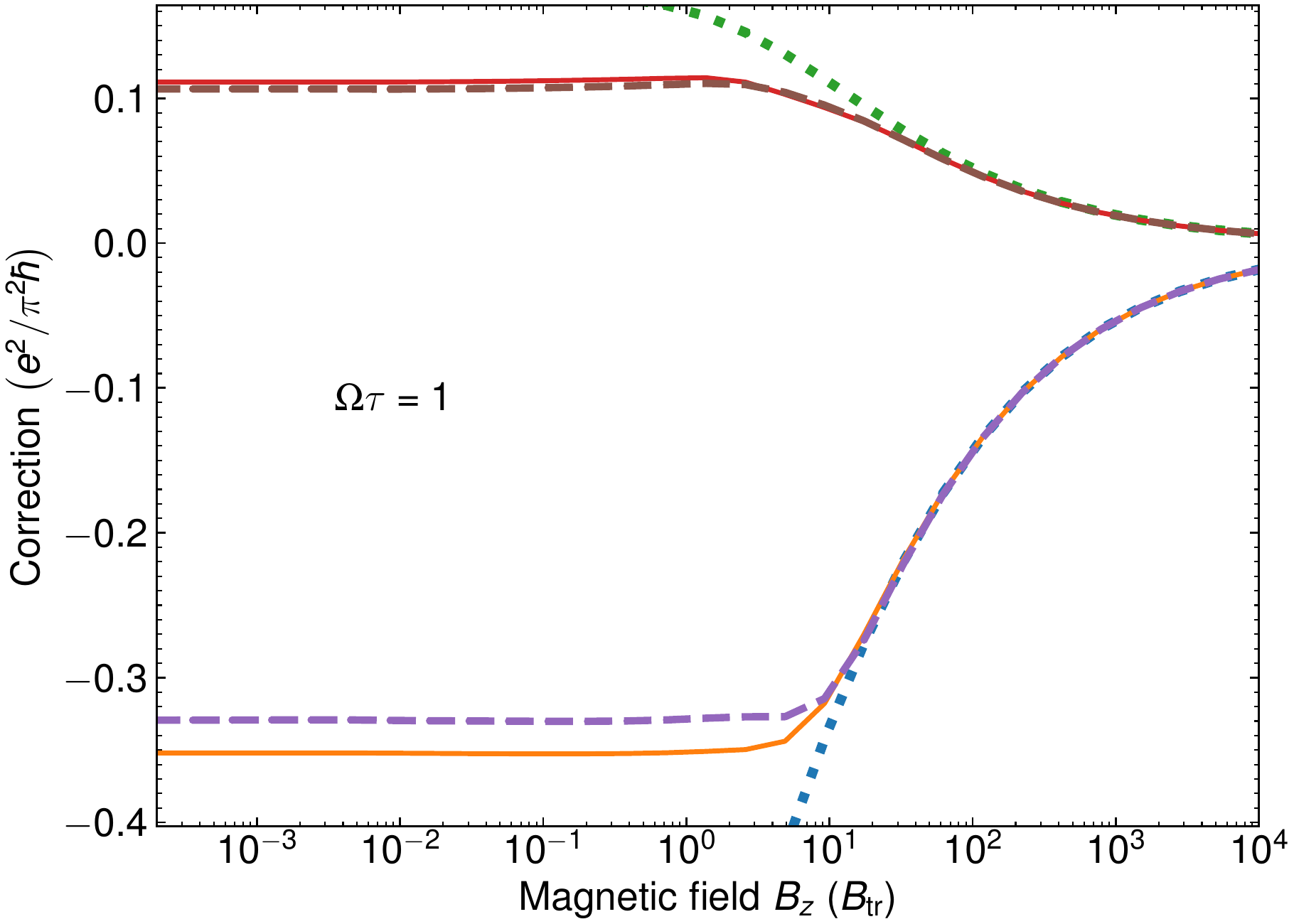}
\caption{Partial contribution to the WL conductivity correction at $\Omega\tau=1$ and $\tau_\phi/\tau=10^3$. The backscattering and non-backscattering contributions for singlet, triplet at $k^2$-splitting and triplet at $k^4$-splitting are shown by dotted, solid and dashed lines, respectively.}
\label{fig_bs_nbs}
\end{figure}

The conductivity at $k^4$-splitting is very close to these dependencies. 
Difference in the magnetoconductivity at $k^2$- and $k^4$-splittings is demonstrated in Fig.~\ref{fig_bs_nbs}. Not only singlet contributions independent of the splitting but also  triplet contributions are very close to each over at two types of splittings. Therefore the results of Fig.~\ref{fig_mcond} are valid for the $k^4$-type of splitting as well. 

\begin{figure}[t]
\centering
\includegraphics[width=0.99\linewidth]{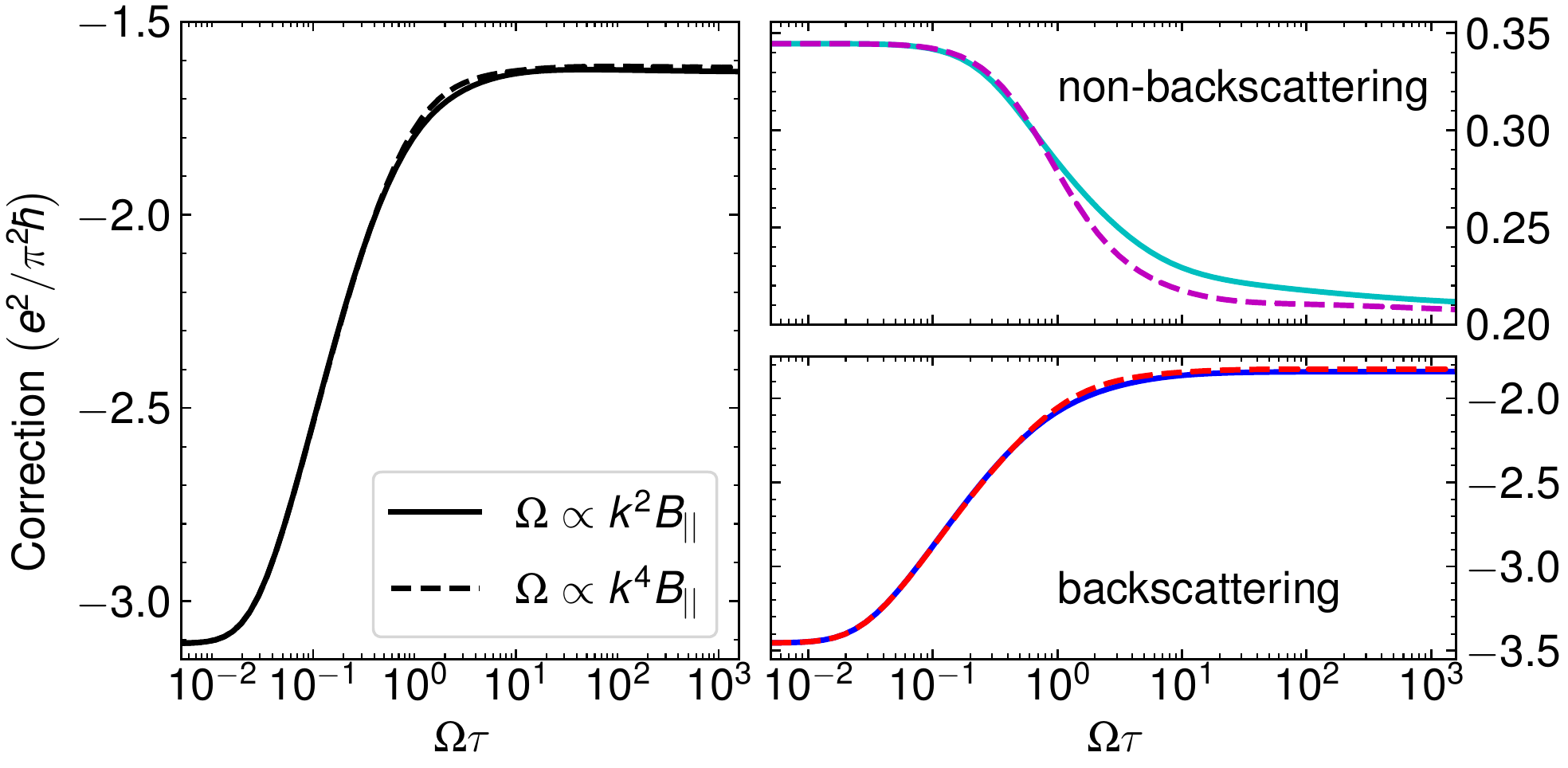}
\caption{Conductivity correction at $B_z=0$ as a function of $\Omega\tau \propto B_\parallel$  at $\tau_\phi/\tau=10^3$.
The total conductivity correction, backscattering and non-backscattering contributions are shown in the left, upper right and lower right panels, respectively. Solid and dashed cureves correspond to the  $k^2$- and $k^4$-splittings.}
\label{fig_zero_field}
\end{figure}

The WL correction to conductivity at $B_z=0$ is analyzed in Fig.~\ref{fig_zero_field}.
According to Eqs.~\eqref{ds0_nondiff},~\eqref{ds0_Omega_infty} the nonbackscattering correction 
at $k^2$-splitting changes from $(1/2)\ln2 = 0.3466$ to 0.209 (in units $e^2/\pi^2\hbar$) when $\Omega\tau$ raises from zero to infinity. 
The backscattering correction changes from $(-1/2)\ln{(\tau_\phi/\tau)}$ to $(-1/4)\ln{(\tau_\phi/\tau)}-0.113$.
At $k^4$-splitting the backscattering correction tends at $\Omega\tau \to \infty$ to $(-1/4)\ln{(\tau_\phi/\tau)}-0.0991$, and the nonbackscattering one to $0.2059$ (in units $e^2/\pi^2\hbar$), cf. Eq.~\eqref{ds0_Omega_infty_k4}.
We see that the size of the WL correction is a little bit larger in the case of $k^2$-splitting. However, the difference is very small.

\section{Summary}

To summarize, the theory of WL of 2D holes in the presence of an in-plane magnetic field is developed. The momentum-dependent Zeeman splitting is taken into account which can be squared or quartic in $k$. The WL conductivity correction, which is negative, is derived for both cases. Calculations show that the results are very close to each other. The $k$-dependent Zeeman splitting suppresses the WL  correction up to factor of two at large splitting. The positive magnetoconductivity in classically-weak perpendicular magnetic fields is calculated for arbitrary values of the Zeeman splitting.
The  developed  is valid for arbitrary values of the product $\Omega\tau$ but the spin splitting $2\hbar \Omega$ assumed much smaller than the Fermi energy. For higher spin splittings, when they are comparable, one should take into account the difference of the Fermi wavevectors in two spin-splitted subbands as it has been done for large Rashba splittings in Ref.~\cite{Large_Rashba_2016}.

\acknowledgments

We gratefully acknowledge the Foundation for the Advancement of Theoretical Physics and Mathematics ``BASIS''. 
The numerical calculations of M.O.N. were supported by the Russian Science Foundation (Project~19-12-00051).
The analytical expressions are derived by L.E.G. with support from the Russian Science Foundation (Project~22-12-00125).

\appendix

\section{Recurrence relations}\label{sec:rec_rel}

We define for $N \geq m$
%
\begin{equation}
\mathcal P_N^{(m)} =\int\limits_0^\infty \dd x \exp\qty(-ax - {x^2\over 2})x^{m}L_{N-m}^{(m)}(x^2).
\end{equation}
At $N=0$ we have $\mathcal P_0^{(0)} = \sqrt{\pi / 2} \exp(a^2/2) \text{Erfc}(a/\sqrt{2})$,
%
%
for $\mathcal P_N^{(0)}$ with $N \geq 1$ we use the following reccurence relations~\cite{MN_NS_mcond_intervalley,Kawabata}
\begin{multline}
N\mathcal P_N^{(0)} = a\delta_{N1} + (N-2)\mathcal P_{N-3}^{(0)} 
\\ + (N-1+a^2) \qty(\mathcal P_{N-2}^{(0)}  - \mathcal P_{N-1}^{(0)})
\end{multline}
and the numerical procedure explained in details in Ref.~\cite{MN_NS_mcond_intervalley}.
For $m\ge 1$, using the recurrence relations for the Laguerre polynomials, we obtain~\cite{footnote2}
\begin{subequations}\begin{align}
\mathcal P_N^{(1)}-\mathcal P_{N-2}^{(1)} & = \delta_{N1}- a (\mathcal P_{N-1}^{(0)}-\mathcal P_{N-2}^{(0)}),
\\
\mathcal P_{N}^{(m\ge2)}-\mathcal P_{N-2}^{(m\ge2)}  & = (m-1)\mathcal P_{N-2}^{(m-2)} \nonumber\\
 &- a \left(\mathcal P_{N-1}^{(m-1)}-\mathcal P_{N-2}^{(m-1)} \right).
\end{align}\end{subequations}

The WL conductivity correction is determined by $P_N$, $Q_N$, $R_{N}^{(m)}$ and $S_{N}^{(m)}$. They are expressed via $\mathcal P_{N}^{(m)}$ as follows:
\begin{equation}
P_N = {l_B\over l} \mathcal P_N^{(0)}\qty(a = {l_B/\tilde{l}}),
\end{equation}
\begin{equation}
Q_{N} = {l_B\over l} {1\over \sqrt{N}}\mathcal P_{N}^{(1)}\qty(a = {l_B/ l}),
\end{equation}
\begin{equation}
R_{N}^{(m)} = {l_B\over l\sqrt{2}} \sqrt{(N-m)!\over N!}\text{Im}\tilde{P}_N^{(m)},
\end{equation}
\begin{equation}
S_{N}^{(m)} = {l_B\over l} 
\sqrt{(N-m)!\over N!} 
{{\mathcal P}_N^{(m)}\qty(a = {l_B/ l})- \text{Re}\tilde{P}_N^{(m)} \over 2},
\end{equation}
where
\begin{equation}
\tilde{P}_N^{(m)} ={\mathcal P}_N^{(m)}\qty[a = (1-2i\Omega\tau){l_B/ l}].
\end{equation}




\begin{thebibliography}{99}
\bibitem{footnote1}The difference in the expression for $\tau_s$ with Ref.~\cite{polaritons} is caused by the difference in the factor 2 in the definition of $\bm \Omega$ in Eq.~\eqref{H}.
\bibitem{footnote2}Note that there are typos in Ref.~\cite{MN_NS_mcond_intervalley}: in Eq.~(A7d) should be $\tilde{P}_1^1 = 1 - P_0^0$ and in (53) there is an extra factor $(-i)^M$.
\end{thebibliography}
\end{document}